\documentclass[smallextended]{svjour3}%
\usepackage{graphicx}
\usepackage{amsmath}
\usepackage{amsfonts}
\usepackage{amssymb}%
\providecommand{\U}[1]{\protect\rule{.1in}{.1in}}
\RequirePackage{fix-cm}
\smartqed
\begin{document}

\title{Low frequency elastic measurements on solid $^{4}$He in Vycor using a
torsional oscillator}
\author{A. D. Fefferman
\and J. R. Beamish
\and A. Haziot
\and S. Balibar }
\date{Received: date / Accepted: date}
\maketitle

\begin{abstract}
Torsional oscillator experiments involving solid $^{4}$He confined in the
nanoscale pores of Vycor glass showed anomalous frequency changes at
temperatures below 200 mK. These were initially attributed to decoupling of
some of the helium's mass from the oscillator, the expected signature of a
supersolid. However, these and similar anomalous effects seen with bulk $^{4}%
$He now appear to be artifacts arising from large shear modulus changes when
mobile dislocations are pinned by $^{3}$He impurities. We have used a
torsional oscillator (TO) technique to directly measure the shear modulus of
the solid $^{4}$He/Vycor system at a frequency (1.2 kHz) comparable to that
used in previous TO experiments. The shear modulus increases gradually as the
TO is cooled from 1 K to 20 mK. We attribute the gradual modulus change to the
freezing out of thermally activated relaxation processes in the solid helium.
The absence of rapid changes below 200 mK is expected since mobile
dislocations could not exist in pores as small as those of Vycor. Our results
support the interpretation of a recent torsional oscillator experiment that
showed no anomaly when elastic effects in bulk helium were eliminated by
ensuring that there were no gaps around the Vycor sample.


\end{abstract}

\institute{A. D. Fefferman \and J. R. Beamish \and A. Haziot \and S. Balibar \at
Laboratoire de Physique Statistique de l'ENS, associ\'{e} au CNRS et aux Universit\'{e}s D. Diderot et P.M. Curie, 24 rue Lhomond 75231 Paris Cedex 05, France \\
\email{balibar@lps.ens.fr}
\and
J. R. Beamish \at
Department of Physics, University of Alberta, Edmonton, Alberta, Canada T6G 2E1
}

\section{Introduction}

Confinement in porous Vycor glass has long been used to study effects of
finite size and disorder on phase transitions in helium. For example, the
melting curve and the lambda line of $^{4}$He are both shifted by confinement
in Vycor's nanoscale pores \cite{Beamish83,Adams84}. The superfluid critical
behavior of thin helium films adsorbed on Vycor has been studied using an
ultrasonic technique \cite{Mulders89} and the torsional oscillator (TO)
technique \cite{Berthold77}. More recently, TO experiments with solid $^{4}$He
in Vycor \cite{Kim04} showed an unexpected frequency increase at temperatures
below 200 mK; similar behavior was subsequently seen for bulk $^{4}$He
\cite{Kim04a}. This was interpreted as evidence of mass decoupling, the
\textquotedblleft non-classical rotational inertia\textquotedblright\ (NCRI)
that would characterize a supersolid phase. However, the shear modulus of
solid $^{4}$He also changes \cite{Day07} below 200 mK, raising the possibility
that the observed TO frequency changes were due to elastic stiffening rather
than to mass decoupling. Recent analyses \cite{Beamish12b,Maris12} have shown
that in most torsional oscillators the observed frequency changes can be
explained by stiffening effects from changes in the shear modulus of the solid
helium. An experiment in which a compound TO was used to study $^{4}$He in
Vycor \cite{Mi12} at two frequencies showed that the behavior was consistent
with elastic effects but not with mass decoupling. When the original Vycor
experiments \cite{Kim04} were recently repeated \cite{Kim12} with the Vycor
coated by epoxy to eliminate any gaps, the TO anomaly disappeared. This
strongly suggests that the anomaly originated in the bulk helium in gaps
between the Vycor and walls. However, since the dimensions of these gaps are
not known for the original oscillator, it has not been possible to confirm
that elastic effects are of the correct magnitude to explain the observed
frequency changes. Elastic changes in the solid $^{4}$He within the Vycor
pores may also contribute to the TO response.

The elastic properties of solid helium confined in Vycor have been studied in
several ultrasonic experiments. The first of these \cite{Beamish83} used 20
MHz shear waves to study the freezing and melting curves of $^{4}$He confined
in Vycor. These measurements were made at temperatures above 300 mK but
subsequent experiments \cite{Beamish91} extended to lower temperatures (100
mK) and covered a range of frequencies (5 to 140 MHz). Freezing was
accompanied by an increase in the shear modulus of the Vycor/helium system of
about 0.7\%, roughly the change expected from the contribution of solid $^{4}%
$He to the system's total shear modulus \cite{Beamish83}. However, the modulus
changes were gradual, beginning at the onset of freezing around 2 K and
extending to temperatures below 0.5 K. The stiffening was also
frequency-dependent, shifting to higher temperatures as the frequency
increased. This behavior was interpreted in terms of stress relaxation in the
confined solid helium via thermally activated vacancy diffusion. Extrapolating
the frequency dependence seen at MHz frequencies to a typical TO frequency (1
kHz) would predict stiffening near 0.5 K.

The apparent discovery of supersolidity in the Vycor/helium system created new
interest in acoustic measurements. Since the transverse sound speed
$v_{t}=\sqrt{\mu/\rho}$ depends on the system's total density $\rho$, as well
as on its shear modulus $\mu$, $v_{t}$ would increase if superfluid decoupled
from the Vycor and reduced the system's effective density \cite{Beamish83}.
Previous measurements extended to 100 mK \cite{Beamish91} but did not show
such an anomaly in the transverse sound speed. However, the TO anomaly began
to disappear above an apparent \textquotedblleft critical
velocity\textquotedblright\ (about 10 $\mu$m/s) and the ultrasonic
measurements were made at higher oscillation speeds. More recently, an
experiment was reported \cite{Kobayashi06} that was designed to search for
supersolidity in Vycor using longitudinal ultrasound at 8 MHz. It extended the
temperature range to 50 mK but, more importantly, used a continuous standing
wave technique to make measurements at velocities as low as 0.1 $\mu$m/s.
Decoupling of 1\% of the solid helium's mass would have been detectable but no
such velocity increase was observed.

If the elastic behavior of solid helium in the Vycor pores is similar to that
of bulk $^{4}$He, much larger sound speed changes would be expected in the
temperature range between 50 and 200 mK. Shear modulus changes as large as
80\% have been seen in low frequency measurements on bulk $^{4}$He
\cite{Haziot13a}. A comparable change for helium in Vycor would change the
system's transverse sound speed by about 0.2\%. However, the shear modulus
changes in bulk $^{4}$He are due to the motion of dislocations which can glide
very easily in the basal plane of hcp $^{4}$He
\cite{Haziot13a,Haziot13b,Haziot13c,Fefferman14}. At high frequencies these
dislocations are strongly damped by thermal phonons, which greatly reduces
their effects on elastic constants. At ultrasonic (MHz) frequencies, velocity
changes due to dislocations in solid $^{4}$He are orders of magnitude smaller,
around 0.2\% \cite{Iwasa79} and the corresponding changes in the Vycor/helium
system's sound speed would be extremely small (less than 10 ppm). At an even
more basic level, we would not expect dislocation motion in solid helium
confined in the pores of Vycor. Even if bulk-like dislocations could exist in
such small pores (diameter around 3 nm), they would be very strongly pinned by
the pore walls. The effects of pore walls on dislocations have been tested
\cite{Rabbani11} by freezing $^{4}$He inside silica aerogel, a much more open
porous material whose silica strands should be effective dislocation pinning
centers. As expected, the low frequency shear modulus did not show large
changes below 200 mK. Instead, it increased gradually over the entire
temperature range between freezing (around 2 K) and the lowest temperatures
(40 mK).

In this paper we report low frequency measurements of the shear modulus of
Vycor whose pores contain solid $^{4}$He. We do not observe any sudden changes
in the temperature range below 200 mK where previous torsional oscillator and
elastic measurements showed anomalies. Instead, we found a small, gradual
increase in the system's shear modulus as it was cooled from 1 K to the lowest
temperature (20 mK). This supports the conclusion from recent Vycor TO
experiments that the low temperature anomalies seen in previous experiments
must have originated in elastic effects in bulk helium in gaps around the
Vycor. Confinement in the small pores of Vycor eliminates the dislocation
movement responsible for the low temperature modulus changes in bulk solid
$^{4}$He. The gradual modulus changes we observe in the present experiments
appear at lower temperatures than the corresponding stiffening seen in
previous ultrasonic measurements at MHz frequencies, as expected for a
thermally activated relaxation process in the solid helium in the pores. The
modulus changes due to the helium, combined with those due to two level
systems (TLS) in the Vycor \cite{Beamish84,Mulders93,Kobayashi06}, would
produce a background temperature dependence in the frequency of a torsional
oscillator containing solid helium-filled Vycor.

\section{Experiment}

The shear modulus of the Vycor/helium system is dominated by that of the Vycor
since the shear modulus of Vycor ($\approx10^{10}$ Pa) is approximately 1000
times greater than that of solid helium. This means that a 1\% change in the
helium's shear modulus only changes the modulus of the system by about a
factor of $10^{-5}$. Ultrasonic velocity measurements can detect relative
changes as small as $10^{-6}$, but it is challenging to achieve comparable
resolution at low frequencies. We would like to have a resolution of order
$10^{-5}$ for our measurements but this is not practical with the direct shear
modulus method we have used for low frequency measurements on bulk $^{4}$He.
High $Q$ torsional oscillators have been highly developed for the study of
superfluidity because of their extremely high mass sensitivity and have been
used to study solid helium for the same reason. However, torsional oscillators
unexpectedly turned out to be quite sensitive to elastic effects, for example
to changes in the shear modulus of solid helium in the hole typically found in
their torsion rod \cite{Beamish12b}. This led us to design a torsional
oscillator in which the Vycor sample was essentially the torsion rod, rather
than the inertial element of the oscillator.

\begin{figure}[ptb]
\includegraphics[width=\textwidth]{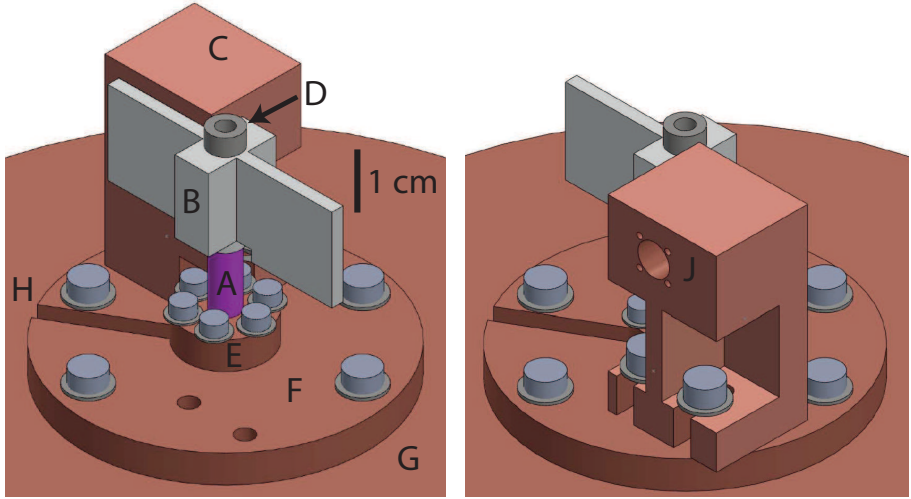}\caption{Opposite views of
the torsional oscillator assembly. The torsional oscillator consists of a
Vycor torsion rod (A) and a magnesium inertial element (B), and its motion is
driven and detected by a fixed electrode (C). The screw (D) is used to press a
grounding wire against the inertial element (not shown). The torsion rod was
glued to a pedestal (E), which was screwed to a bottom plate (F), which was in
turn screwed to a massive vibration isolation plate (G). A slot (H) was cut in
the bottom plate to make room for the fill line (not shown). The back of the
fixed electrode is shown in the opposite view, with a hole for the MCX
electrical connector (J).}%
\label{fig:TOdraw}%
\end{figure}

Figure \ref{fig:TOdraw} shows a model of the torsional oscillator assembly.
The torsion rod consists of a 3.45 mm diameter Vycor porous glass cylinder
coated by a 0.5 mm thickness of Stycast 1266 epoxy. The Vycor is about 30\%
porous with a typical pore diameter of 4 nm. The bottom of the Vycor cylinder
was first glued into a socket formed by a narrow lip on top of the copper
pedestal labeled in Fig. \ref{fig:TOdraw} using Stycast 2850. A jig was used
to keep the Vycor cylinder aligned with the axis of the copper pedestal as the
Stycast 2850 cured. The Vycor was then potted in Stycast 1266 out to the
diameter of the pedestal. The Stycast 1266 was then machined to a 4.5 mm
diameter, leaving the end and the side of the Vycor cylinder sealed by epoxy.
The copper pedestal had an axial hole that was connected to a CuNi capillary
on the bottom so that the Vycor could be pressurized with helium. At room
temperature, we observed diffusion of helium through the epoxy layer, but the
diffusion signal vanished at low temperatures, indicating that the fill line
was open and the Vycor was sealed. The magnesium inertial element was glued
onto the torsion rod with Stycast 1266. Since the screws that fastened the
fixed electrode went into slots, there was some freedom in the orientation of
the fixed electrode, and it was only necessary to approximately orient the
inertial element on the torsion rod before the epoxy cured. One end of a 0.2
mm diameter copper wire was screwed to the top of the inertial element and the
other end was screwed to the copper bottom plate (Fig. \ref{fig:TOdraw}). This
wire was used to ground the inertial element thermally and electrically. The
geometry of the inertial element and the density of magnesium (1.74 g/cc)
imply that the moment of inertia of the torsional oscillator was $I=3.03$ g
cm$^{2}$ (the moment of inertia of the torsion rod is only 0.01 g cm$^{2}$).
The low temperature shear moduli of Vycor and Stycast 1266 are respectively
7.0 \cite{Watson03} and 2.0 GPa \cite{Pohl02}, so that their contributions to
the torsion constant $\kappa_{0}=18$ N\thinspace m of the composite torsion
rod were respectively 12 N\thinspace m and 6.2 N\thinspace m. Thus the
resonant frequency is expected to be 1200 Hz.

The bottom plate of the torsional oscillator assembly was screwed to a 1 kg
copper vibration isolation plate (Fig. \ref{fig:TOdraw}). This vibration
isolation plate was suspended from the refrigerator by a long, thin copper
rod, forming a second torsional oscillator with a torsional resonance at about
10 Hz. Thus vibrations of the TO\ were in principle attenuated by the square
of the ratio of the frequencies of the torsional oscillators, i.e., a factor
of 10,000. In addition, the entire refrigerator was floating on air legs, thus
further decreasing vibrations of the TO.

\begin{figure}[ptb]
\includegraphics[width=0.75\textwidth]{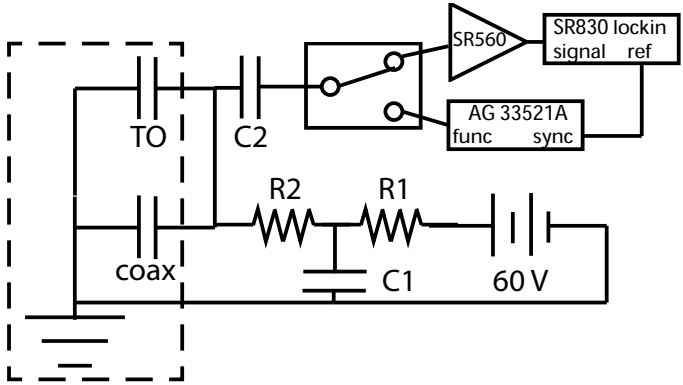}\caption{Circuit for
excitation of the torsional oscillator and measurement of its motion during
its free decay. The Keithley 7001 switch is used to select drive or detect
modes, and the dashed box indicates the cryogenic portion of the circuit. The
capacitor labeled \textquotedblleft TO\textquotedblright\ represents the 8.0
pF capacitance between the inertial element and the fixed electrode, which
varies as the TO oscillates. The capacitor labeled \textquotedblleft
coax\textquotedblright\ is the $\approx$300 pF capacitance of the coaxial
cable that connects the TO to the measurement system. C1 and C2 are 3 nF
capacitors and R1 and R2 are 10 M$\Omega$ resistors. An Agilent 33521A
function generator was used to drive the motion, and a Stanford Research 560
voltage amplifier and a Stanford Research 830 lockin amplifier were used to
detect the motion.}%
\label{fig:circuit}%
\end{figure}

Figure \ref{fig:circuit} shows the circuit used to drive and detect the motion
of the TO. The resonant frequency of the\ TO\ was inferred from free decay
measurements, and the same fixed electrode was used to drive the TO and
measure the decay of its amplitude of motion. In both drive and detect modes,
a bias voltage $V_{0}=60$ V was applied across the capacitor formed by the
inertial element of the TO and the fixed electrode. In drive mode, a signal
near the resonant frequency of the TO with a magnitude of order 1 V was added
to the voltage across this capacitor by switching to the function generator
(Fig. \ref{fig:circuit}). Thus the TO was driven by a force proportional to
the product of the dc and ac components of the drive voltage. After the TO
reached its equilibrium response amplitude, the switch was set to the voltage
preamplifier.\ Due to the high input impedance of the preamplifier, the charge
on the capacitor labeled \textquotedblleft TO\textquotedblright\ remained
approximately constant on the time scale of the TO oscillation period. Thus
the oscillation amplitude of the TO is given by $d=VC_{\text{coax}}d_{0}%
/V_{0}C_{\text{TO}}$, where $V$ is the voltage at the input of the
preamplifier, $d_{0}=0.3$ mm is the equilibrium spacing between the inertial
element and the fixed electrode, and $C_{\text{coax}}\approx300$ pF and
$C_{\text{TO}}\approx8$ pF are respectively the capacitances labeled
\textquotedblleft coax\textquotedblright\ and \textquotedblleft
TO\textquotedblright\ in Fig. \ref{fig:circuit}. In detect mode, the reference
frequency of the lockin amplifier was set close to the resonant frequency of
the TO.

\begin{figure}[ptb]
\includegraphics[width=\textwidth]{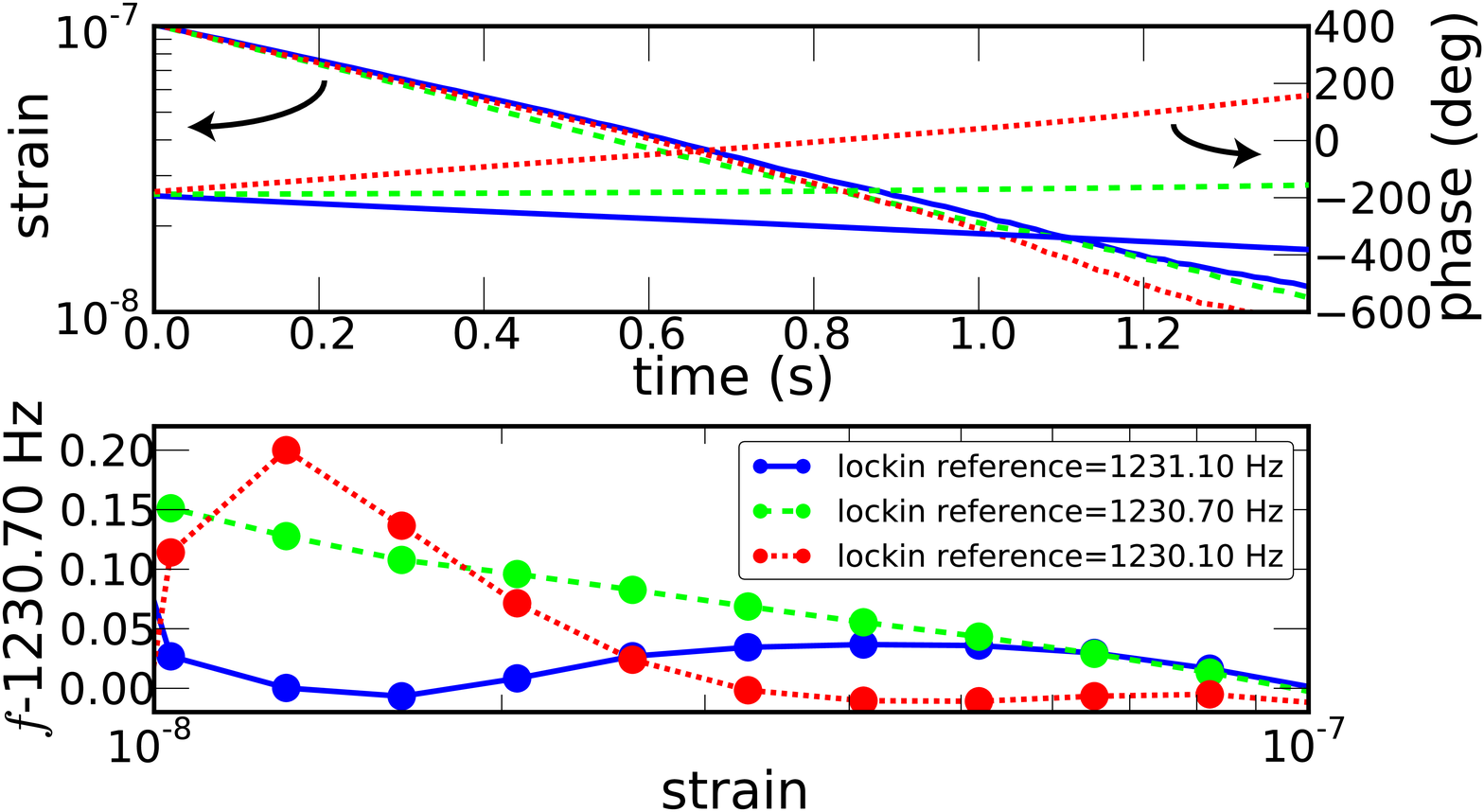}\caption{Free decay of
torsional oscillations with solid-helium filled Vycor pores at 20 mK. \ The
measurements were made with the lockin reference frequency at the resonant
frequency $f=1230.7$ Hz (green) and near the half-power points of the
resonance (blue and red). Upper panel: Magnitude of the maximum shear strain
in the Vycor rod and its phase relative to the lockin reference signals at the
three different frequencies. Lower panel: Resonant frequency $f$ of the
torsional mode as a function of strain, obtained by averaging the time
derivative of the phase over many free decays. This panel shows that the value
of $f$ that we obtain is independent of strain as well as the reference
frequency.}%
\label{fig:decay}%
\end{figure}

The maximum shear strain in the Vycor is given by $d\left(  D_{\text{vy}%
}/D_{\text{wing}}L\right)  $, where $D_{\text{vy}}=3.45$ mm is the diameter of
the Vycor, $D_{\text{wing}}=23$ mm is the distance between the midpoints of
the two fixed electrodes, and $L=8.5$ mm is the length of the torsion rod. The
top panel of Fig. \ref{fig:decay} shows three examples of the magnitude of the
maximum shear strain in the Vycor and its phase relative to the lockin
reference frequency during the free decay. Those three measurements were made
at 20 mK and with solid helium-filled Vycor pores (crystal X1, see below). The
time derivative of the strain yields a TO quality factor $Q\approx1500$. The
three different colors correspond to different lockin reference frequencies
within the width of the torsional oscillator resonance $f/Q\approx1$ Hz. The
difference between the oscillation frequency of the TO and the lockin
reference is given by the time derivative of the phase of the shear strain.
Thus we can determine the resonant frequency of the TO as a function of time,
and thus as a function of strain, since the TO oscillates at the resonant
frequency during free decay. The bottom panel of Fig. \ref{fig:decay} shows
the resonant frequency of the TO as a function of strain. These data points
were determined from the average of many repetitions of the measurement shown
in the top panel. The fact that the measured resonant frequency is nearly
independent of the lock-in reference frequency demonstrates the reliability of
this measurement technique, and the absence of strain dependence of the
resonant frequency demonstrates that nonlinear effects are negligible in this
strain range.

For this experiment, we used natural purity helium, which has a nominal $^{3}%
$He concentration of $3 \times10^{-7}$. The solid samples were formed using
the blocked capillary technique, so that their growth occurred at nearly
constant volume. The volume of the fill line connected to the TO was
relatively large, and it was significant compared with the volume of the pores
in the Vycor. Thus the growth technique was chosen to minimize the amount of
liquid removed from the Vycor due to freezing of the bulk helium in the fill
line, which occurred first. After ceasing to pump on the 1 K\ pot in order to
grow a new solid sample, the TO was allowed to warm to a target temperature.
During this process, the \textquotedblleft1 K\textquotedblright\ pot reached
temperatures as high as 3 K. The helium pressure was set low enough so that
the helium was liquid in the pores and the fill line. We assumed that at this
point the temperature along the fill line increased from the TO to the 1 K
pot. Thus when the pressure in the fill line was subsequently increased with
the TO at the target temperature, the solid plug should have formed in the
fill line close to the TO, where the temperature and consequently the melting
pressure were lowest. Having the plug form close to the TO minimized the
pressure decrease in the Vycor pores as the rest of the helium in the fill
line solidified after restarting the 1 K\ pot.

\section{Results and Discussion}

\begin{figure}[ptb]
\includegraphics[width=0.5\textwidth]{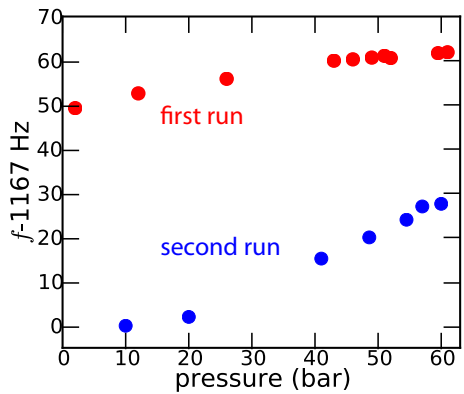}\caption{The resonant
frequency of the torsional oscillator as a function of the pressure of liquid
helium in the Vycor pores measured at temperatures near 3 K.}%
\label{fig:pdepend}%
\end{figure}

Figure \ref{fig:pdepend} shows the dependence of the resonant frequency of the
TO on the pressure inside the Vycor. In between the two sets of measurements
presented in this paper (denoted as \textquotedblleft first
run\textquotedblright\ and \textquotedblleft second run\textquotedblright),
the TO was warmed to room temperature. In the second run, the dependence of
the resonance on pressure was larger ($\approx$5 bar/Hz) (Fig.
\ref{fig:pdepend}). The pressure dependence was measured at temperatures near
3 K with liquid helium filling the Vycor pores. This effect cannot be due to
an increase in the moment of inertia of the torsion rod due to an increase of
its diameter under pressure because that would cause the resonant frequency to
decrease with pressure, which is the opposite of the behavior in Fig.
\ref{fig:pdepend}. It is likely that the epoxy bulged out and partially
detached from the Vycor in response to the pressure, while still maintaining
the seal, thus increasing the diameter of this epoxy shell and consequently
increasing the torsion constant. The decrease of the resonant frequency of the
TO in the second run relative to the first run may have been caused by
differential thermal expansion of the Vycor and epoxy, which could have
produced cracks in the epoxy that softened it. This softening of the epoxy
would also have caused the epoxy to bulge out more in response to pressure,
producing the higher pressure sensitivity of the resonant frequency that was
observed in the second run (Fig. \ref{fig:pdepend}).

The pressure decrease due to solidification of the helium in the Vycor and the
fill line causes a decrease in the resonant frequency of the TO as large as 8
Hz between the bulk melting temperature and 1 K. However, if we assume that
solidification of the helium in the Vycor pores is complete by the time the
TO\ temperature reaches 1 K, then the change in the resonant frequency of the
TO due to pressure changes should be negligible below this temperature. The
pressure of bulk helium at constant volume decreases by less than 50 mbar as
it is cooled from 1 K to 0 K\cite{Jarvis68}. If we assume that the pressure
change of helium in Vycor is similar to that of bulk helium, then a 50 mbar
pressure decrease should cause a TO frequency \textit{decrease} of only 10
mHz, which is much less than the measured frequency increase between 0 and 1 K
(see below).

Figure \ref{fig:fvsT} shows the resonant frequency of the TO as a function of
temperature for empty, liquid helium-filled and solid helium-filled pores. Due
to the pressure dependence of the resonant frequency discussed above, there
were significant offsets in the resonant frequencies measured at different
pressures. In Fig. \ref{fig:fvsT}, the curves are shifted so that they overlap
at 15 mK, and the offset frequency $f_{0}$ for each curve is noted in the
legend, along with the pressure in the fill line measured at room temperature
just before it was blocked by solid helium. The measurements with empty and
liquid-filled pores show the background contribution of the amorphous
materials (Vycor and epoxy) used to construct the torsion rod. The temperature
dependence is probably due to a broad distribution of tunneling two level
systems (TLS). The TLS cause a maximum in the stiffness of amorphous materials
at a temperature that scales as the cube root of the drive
frequency\cite{Fefferman10}. The resonant frequency of the TO is sufficiently
small so that this maximum is at or below the base temperature of our
cryostat, so that the background contribution to the resonant frequency
decreases with temperature. The small difference between the measurements with
an empty cell and with liquid-filled pores could be due to the effect of the
liquid helium on the relaxation of the TLS\cite{Mulders93}. The measurements
shown in Fig. \ref{fig:fvsT} with solid helium-filled pores show the combined
effects of the TLS in the amorphous materials and the solid helium in the pores.

The TO has a $Q\approx1500$ below 1 K, with a slight increase at the lowest
temperatures. This is consistent with the nearly universal and temperature
independent $Q$ of amorphous solids that is observed above the temperature of
the stiffness maximum and below a few kelvin \cite{Pohl02}. If a process with
a single relaxation time were responsible for the TO frequency shift caused by
the solid helium in the pores (Fig. \ref{fig:fvsT}), then a dissipation peak
should appear near the temperature of the midpoint of the frequency shift in
the case of solid-filled pores. The magnitude of this dissipation peak should
be of order the frequency shift \cite{Nowick72}, i.e., of order $10^{-3}$ in
this experiment (Fig. \ref{fig:fvsT}). The absence of such a dissipation peak
suggests that the relaxation process in the solid helium has a distribution of
relaxation times, so that the process causes a smaller dissipation peak for a
given frequency shift.

\begin{figure}[ptb]
\includegraphics[width=\textwidth]{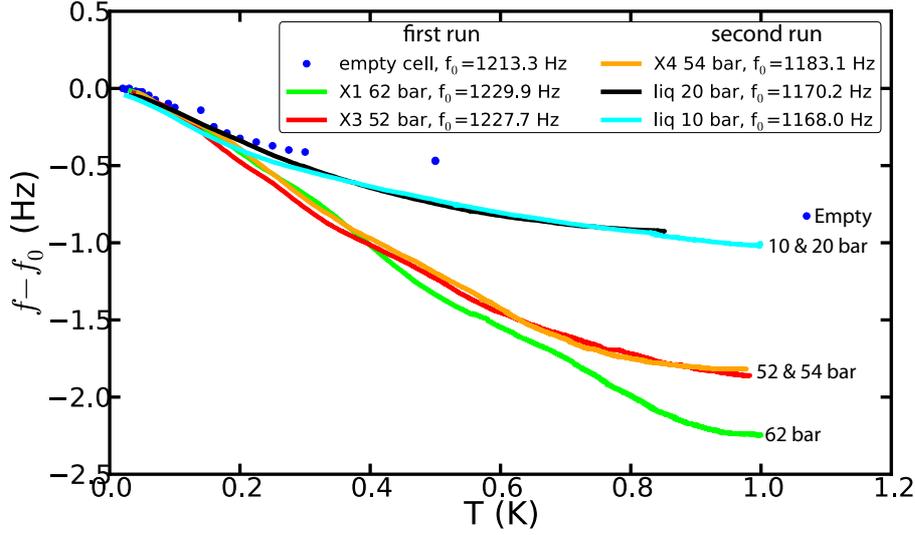}\caption{The resonant
frequency $f$ versus the temperature $T$ with solid-filled, liquid-filled or
empty Vycor pores at a strain of 10$^{-7}$. The measurements were made on
warming. For the solid samples, the pressure before the helium in the fill
line froze is indicated in the legend. The $T$ dependence of $f$ with
liquid-filled or empty pores is due to the amorphous solids composing the
torsion rod. With solid-filled Vycor, vacancy diffusion in the helium also
contributes to the $T$ dependence. The variation of the offset frequencies
$f_{0}$ is mainly due to the pressure dependence of the resonance (Fig.
\ref{fig:pdepend}.) }%
\label{fig:fvsT}%
\end{figure}

An increase in the shear modulus of the helium filling the Vycor pores will
cause the shear modulus of the helium+Vycor system to increase. However, we
emphasize that the change in the shear modulus of the helium+Vycor is not
equivalent to the change in the shear modulus of the helium. Instead, the
change in the shear modulus of the helium+Vycor should be of the order of the
porosity of the Vycor times the change in the shear modulus of the helium
\cite{Maris12}. The change in the contribution of the helium to the shear
modulus of the torsion rod can be determined from the measurements of the
resonant frequency with solid-filled pores (Fig. \ref{fig:fvsT}) by
subtracting the background contribution. The resonant frequency of the
torsional oscillator is%
\begin{equation}
f=\frac{1}{2\pi}\sqrt{\frac{\kappa}{I}}%
\end{equation}
where $\kappa$ is the torsion constant. Since the moment of inertia is
dominated by the magnesium inertial element, and thus is independent of
temperature, the relative change in the resonant frequency is%
\begin{equation}
\frac{\delta f}{f_{0}}=\frac{1}{2}\frac{\kappa-\kappa_{0}}{\kappa_{0}}
\label{eq:rel_freq_shift}%
\end{equation}
where $\delta f\left(  T,p\right)  =$ $f-f_{0}=f-f\left(  T=15\text{
mK},p\right)  $ is the quantity plotted in Fig. \ref{fig:fvsT}, $p$ is the
pressure, and $\kappa_{0}=\kappa\left(  T=15\text{ mK},p\right)  $. We have%
\begin{equation}
\kappa\left(  T,p\right)  =\kappa_{0}+\Delta\kappa_{\text{TLS}}+\frac{\pi
D_{\text{vy}}^{4}}{32L}\Delta\mu_{\text{He}} \label{eq:kappa}%
\end{equation}
where $\Delta\kappa_{\text{TLS}}\left(  T\right)  $ is the change in the
torsion constant due to two level systems relative to $\kappa_{0}$, and
$\Delta\mu_{\text{He}}\left(  T,p\right)  $ is the change in the contribution
of the helium to the shear modulus relative to its contribution at 15 mK and
pressure $p$. As explained above, the contribution of the helium to the shear
modulus of the torsion rod is distinct from the shear modulus of the helium
itself. With liquid filled pores at 10 bar, $\Delta\mu_{\text{He}}=0$, so that
Eqs. \ref{eq:rel_freq_shift} and \ref{eq:kappa} yield%
\begin{equation}
\Delta\kappa_{TLS}=\frac{2\kappa_{0}}{f_{0}}\delta f\left(  T,10\text{
bar}\right)  \label{eq:dkappaTLS}%
\end{equation}
and with solid-filled pores at pressure $p_{s}$ Eq. \ref{eq:rel_freq_shift}
yields%
\begin{equation}
\kappa-\kappa_{0}=\frac{2\kappa_{0}}{f_{0}}\delta f\left(  T,p_{s}\right)
\label{eq:dkappaHe}%
\end{equation}
where $\kappa_{0}/f_{0}=0.015$ N\thinspace m/Hz was calculated from the values
of $I$ and $\kappa_{0}$ given above. Rearranging Eq. \ref{eq:kappa} and
substituting Eqs. \ref{eq:dkappaTLS} and \ref{eq:dkappaHe}, we obtain%
\begin{equation}
\Delta\mu_{He}=\frac{64L\kappa_{0}}{\pi D_{\text{vy}}^{4}f_{0}}\left[  \delta
f\left(  T,p_{s}\right)  -\delta f\left(  T,10\text{ bar}\right)  \right]
\label{eq:dmu}%
\end{equation}

\begin{figure}[ptb]
\includegraphics[width=0.75\textwidth]{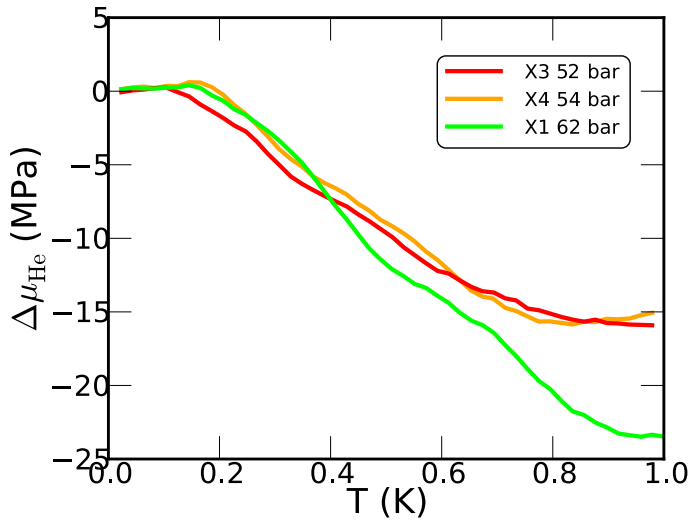}\caption{The contribution of
the solid helium to the shear modulus of the torsion rod versus temperature
for the three solid samples of Fig. \ref{fig:fvsT}, determined using Eq.
\ref{eq:dmu}.}%
\label{fig:muvsT}%
\end{figure}

The temperature dependence of $\Delta\mu_{He}$ for the three solid samples in
Fig. \ref{fig:fvsT} is shown in Fig. \ref{fig:muvsT}. Below $\approx150$ mK,
$\Delta\mu_{He}$ is temperature independent within 0.5 MPa. As the temperature
is increased above 150 mK, $\Delta\mu_{He}$ gradually decreases and levels off
at $\approx900$ mK. The magnitude of the softening over the entire temperature
range increases with the initial pressure in the fill line, as expected since
the shear modulus of solid helium increases with pressure \cite{Beamish12a}.
The temperature dependence in Fig. \ref{fig:muvsT} is qualitatively different
from that of the shear modulus of bulk $^{4}$He with 300 ppb $^{3}$He at
$\approx1$ kHz. Indeed, in the bulk case, there is a relatively sudden
decrease in the shear modulus near 100 mK due to thermal unbinding of $^{3}$He
atoms from dislocations, above which the shear modulus is nearly temperature
independent up to 1 K\cite{Haziot13c}. Thus the softening in Fig.
\ref{fig:muvsT} is not due to dislocations, as expected since the pores of
Vycor are too small to accommodate dislocations.

The softening observed in ultrasonic measurements of the shear modulus of
helium-filled Vycor \cite{Beamish91} was explained in terms of vacancy
diffusion, and we think the same physics is responsible for the softening in
Fig. \ref{fig:muvsT}. Vacancy flow relaxes stress gradients by transporting
mass. The relaxation time for a crystal of size $L$ is \cite{Beamish91}%
\begin{equation}
\tau=\frac{k_{B}TL^{2}}{2\alpha\mu VD}\label{eq:tau}%
\end{equation}
where $\alpha$ is a shape dependent numerical factor, $\mu$ is the shear
modulus, and $V$ is the vacancy volume. The self-diffusion coefficient of
solid helium is given by%
\[
D=D_{0}e^{-E_{V}/k_{B}T}%
\]
where $E_{V}$ is the activation energy for vacancy diffusion. For $\omega
\tau<<1$, there is sufficient time for vacancies to relax the stress on each
strain cycle, whereas for $\omega\tau>>1$, there is not. Here, $\omega=2\pi
f$. Thus we expect the midpoint of the stiffening transition to occur near a
temperature $T_{\text{max}}$ such that $\omega\tau=1$:%
\begin{equation}
\omega T_{\text{max}}=\left[  \frac{\alpha\mu VD_{0}}{k_{B}L^{2}}\right]
e^{-E_{V}/k_{B}T_{\text{max}}}\label{eq:wTmax}%
\end{equation}
In \cite{Beamish91}, $E_{V}$ for helium in Vycor was determined from the
$\omega$ dependence of the attenuation peak that occurs at $T_{\text{max}}$.
In \cite{Molz_thesis}, the same technique was used to show that $E_{V}$ is
21.9 K, 17.7 K and 13.6 K for pressures of 81.6 bar, 55.8 bar and 43.6 bar,
respectively. Thus $E_{V}$ increases with pressure, and the temperature of the
midpoint of the stiffening transition increases with initial pressure in Fig.
\ref{fig:muvsT}, as expected from Eq. \ref{eq:wTmax}. Since $\Delta\mu_{He}$
could only be measured at a single frequency in the present work, and the
pressure in the Vycor is not precisely known, $E_{V}$ could not be determined.
However, the pressure of the solid in the Vycor for the present measurements
is between the maximum pressure in the fill line before it was plugged (i.e.
62 bar) and the minimum freezing pressure in Vycor ($\approx40$
bar\cite{Beamish83}). If we make a linear fit to the pressure dependence of
$E_{V}$ observed in \cite{Molz_thesis}, then $E_{V}$ in the present work would
be between 13.3 and 18.1 K. Following Ref. \cite{Beamish91}, we take
$\alpha=16$, $V=3.2\times10^{-29}$ m$^{3}$, $\mu=1.9\times10^{7}$ Pa,
$L=3.1\times10^{-9}$ m and $D_{0}=1.4\times10^{-8}$ m$^{2}$/s. For $f=1200$
Hz, Eq. \ref{eq:wTmax} then implies that $T_{\text{max}}$ is between 0.67 K
and 0.93 K. (The calculated value of $T_{\text{max}}$ depends weakly on the
value of $\mu$.) This range in $T_{\text{max}}$ is at the upper end of the
range of temperatures over which $\Delta\mu_{\text{He}}$ varies in Fig.
\ref{fig:muvsT}. However, for a Debye relaxation process, $T_{\text{max}}$
should be at the midpoint of the stiffening transition \cite{Nowick72}. This
discrepancy could have the same explanation as the absence of a resolvable
dissipation peak at the midpoint of the stiffening transition with
solid-filled pores, i.e., a distribution of relaxation times. The behavior
observed in Ref. \cite{Beamish91} also could not be explained by a Debye
relaxation process with a single activation energy.

The torsion constant of our TO was not significantly affected by the bulk
helium region between the Vycor and its epoxy layer that would have existed if
the epoxy partially detached from the Vycor in response to pressure (see
above). Experimentally, this is clear from the observed temperature
dependence: we did not observe a sharp increase in the torsion constant near
100 mK due to $^{3}$He pinning of dislocations in the buk region. This is not
surprising because such a bulk
region would be small enough so that a change in its shear modulus would not
greatly affect our results. A 120 $\mu$m radial expansion of the epoxy shell is
required to account for the $\approx5\%$ change in the torsion constant
between 0 and 60 bar in the second run (Fig. \ref{fig:pdepend}). To simplify
our calculation of the expansion, we have neglected the probable increase in
the shear modulus of the epoxy layer and the decrease in its thickness after
expansion. Below 1 K, the pressure of the 120 $\mu$m thick bulk $^{4}$He region would be
less than 60 bar, so that its intrinsic shear modulus would be less than 20 MPa. If
we assume that dislocation motion causes the shear modulus of the bulk region
to vanish at 1 K, then the maximum corresponding decrease in the torsion constant would
be 0.05\%. In fact, a smaller decrease is expected, since the decrease in the
shear modulus for polycrystals is typically smaller than in single crystals
and is sometimes as small as 10\% \cite{Day07}. In comparison, the maximum observed
change in the torsion constant between 0 and 1 K was 0.4\% (Fig. \ref{fig:fvsT}).

\begin{figure}[ptb]
\includegraphics[width=0.75\textwidth]{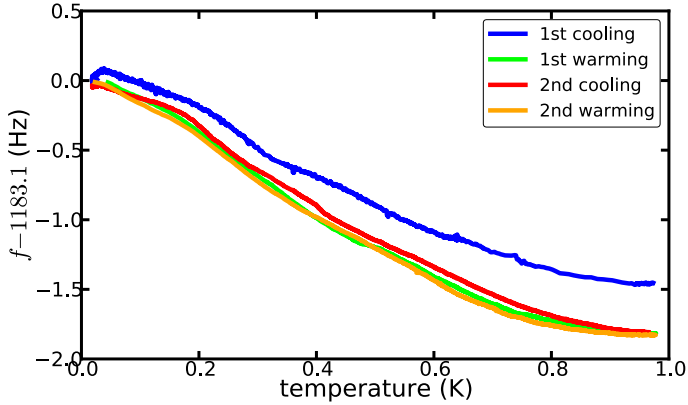}\caption{The resonant
frequency $f$ versus temperature for two thermal cycles of sample X4. As in
the other samples, a slow relaxation process occurred during the first cooling
of each sample, after which the measurement was reproducible upon thermal
cycling. The measurement was made at a strain of 10$^{-7}$.}%
\label{fig:reprod}%
\end{figure}

Figure \ref{fig:reprod} shows the behavior of the resonant frequency of the TO
upon thermal cycling between 0 and 1 K. The curve labeled \textquotedblleft1st
cooling\textquotedblright\ was obtained just after freezing the helium and is
different from the results of the subsequent measurements. After the first
cooling, the results were reproducible. This suggests that a slow relaxation
process occurred during the first cooling. It seems unlikely that this is
relaxation of pressure gradients on the length scale of the torsion rod by
vacancy diffusion because the corresponding stress relaxation time increases
exponentially as temperature is decreased, and yet significant relaxation
continues down to the base temperature, as demonstrated by the difference
between the blue curve and the others at low temperatures in Fig.
\ref{fig:reprod}. Using Eq. \ref{eq:tau} with $E_{V}=13.3$ K, $L=1$ mm and
with the other parameters taking the values from \cite{Beamish91} given above,
we obtain $\tau=8$ days at $T=0.8$ K. The measurements shown in Figs.
\ref{fig:fvsT} and \ref{fig:muvsT} were made on warming, so that the results
were reproducible.

In \cite{Kim12}, the TO frequency shifts that occurred as the helium in the
Vycor froze and with solid helium in the Vycor were attributed to the
stiffening of the Vycor/helium composite. However, it is not clear how such
large frequency shifts could be produced by this stiffening. We can calculate
the expected frequency shift of the TO in \cite{Kim12} using the technique
from \cite{Reppy12}: We take the thick part of the torsion rod, the Invar
plate and the wings to be a perfectly rigid bob.\ The wings might not be
perfectly rigid, but that should be a small, temperature independent
perturbation to the resonant frequency. Then one calculates the torque
$\tau_{v}$ on the bob due to the helium-filled Vycor, accounting for the
finite shear modulus of the Vycor. Inserting $\tau_{v}$ into the equation of
motion for the bob, one can obtain a first order solution for the resonant
frequency as a function of the shear modulus of the Vycor+helium. The period
shift $dp$ of the TO is then related to a change in the shear modulus of the
Vycor+helium $d\mu$ as%
\begin{equation}
dp=-\frac{2\pi^{2}}{3}\frac{I_{v}}{I_{b}}\frac{h^{2}\rho}{p\mu_{\text{V}}^{2}%
}d\mu\label{eq:dp}%
\end{equation}
where $h=10$ mm is the height of the Vycor cylinder, $\rho=$ 1.6 g/cc is the
density of Vycor porous glass, $p=1.15$ msec is the oscillation period of the
TO, $\mu_{\text{V}}=7$ GPa \cite{Watson03} is the shear modulus of Vycor
porous glass and $I_{b}=170$ g cm$^{2}$ \cite{Kimper} and $I_{v}=60$ g
cm$^{2}$ are respective estimates of the moments of inertia of the bob and
Vycor assuming they are perfectly rigid. Eq. \ref{eq:dp} implies that the $20$
MPa change in the shear modulus of the Vycor+helium observed below 1 K in the
present work (Fig. \ref{fig:muvsT}) corresponds to a period shift of $dp=130$
psec, but a period shift of 900 ps relative to the empty cell was observed
between 30 and 900 mK in \cite{Kim12}. The discrepancy between the apparent
and expected changes in the shear modulus in \cite{Kim12} is even larger at
higher temperatures. Eq. \ref{eq:dp} implies that the period shift exceeding 5
nsec observed during freezing and melting of the helium in the Vycor in
\cite{Kim12} corresponds to an apparent shear modulus change of 770 MPa. The
contribution of the helium to the shear modulus of the Vycor+helium system
should be of order the stiffness of the helium \cite{Beamish83}, but 770 MPa
is almost ten times greater than $c_{33}$, the largest component of helium's
stiffness tensor\cite{Beamish12a}, at the 65 bar maximum pressure of the
measurement in \cite{Kim12}. It is more likely that the period increases in
\cite{Kim12} mainly result from bulging of the Vycor portion of the torsional
oscillator and consequent increases in the moment of inertia due to increases
in helium pressure. A 10 nsec shift in the period of the torsional oscillator
corresponds to a relative shift in the total moment of inertia of the
torsional oscillator of 18 ppm, or a relative increase in the moment of
inertia of the Vycor portion by 69 ppm. This in turn corresponds to an
increase in the radius of the Vycor by 17 ppm or 120 nm. It is difficult to
compare this with an estimate of the deformation of the Vycor in response to
changes in pressure because of the porous geometry.

In conclusion, we have shown that resonant frequency shifts observed in previous torsional
oscillator measurements on solid $^{4}$He in Vycor \cite{Kim04,Kim12} were not
mainly due to changes in the shear modulus of the Vycor+helium system
$\Delta\mu_{\text{He}}$. In the case of \cite{Kim04}, the sudden frequency shift
that was observed near 100 mK is inconsistent with the temperature dependence
of $\Delta\mu_{\text{He}}$. We found that $\Delta\mu_{\text{He}}$ is nearly temperature
independent below 150 mK and slowly
decreases as the temperature is increased from 200 mK to 1 K. Thus the sudden
frequency shift near 100 mK in \cite{Kim04} must have been due to an elastic effect
involving a layer of bulk solid helium between the Vycor and the cell wall, as argued
in \cite{Kim12}. In the case of the torsional oscillator used in \cite{Kim12},
the frequency shifts measured during freezing of the helium in
the Vycor and at lower temperatures are too large to have been caused by
changes in the shear modulus of the helium-filled Vycor torsion bob.
Finally, we believe that the explanation for the temperature dependence of
$\Delta\mu_{\text{He}}$ is the freezing of a thermally activated relaxation
process, probably helium vacancy diffusion.

\begin{acknowledgements}
This work was supported by grants from ERC (AdG
247258-SUPERSOLID) and from NSERC Canada.
\end{acknowledgements}


\end{document}